\documentclass{article}
\usepackage{amssymb}
\usepackage{amsfonts}
\usepackage{amsmath}

\input{tcilatex}
\begin{document}

\title{Einstein gravity with generalized cosmological term from
five-dimensional AdS-Maxwell-Chern-Simons gravity }
\author{L. Avil\'{e}s$^{1}$, J. D\'{\i}az$^{1}$, D.M. Pe\~{n}afiel$^{1}$,
V.C. Orozco$^{2}$ and P. Salgado$^{1}$ \\
$^{1}$Instituto de Ciencias Exactas y Naturales, Facultad de Ciencias,\\
Univesidad Arturo Prat, Avda. Arturo Prat 2120, Iquique, Chile.\\
$^{2}$Departamento de F\'{\i}sica, Universidad de Concepci\'{o}n, \\
Casilla 160-C, Concepci\'{o}n, Chile}
\maketitle

\begin{abstract}
Some time ago, the standard geometric framework of Einstein gravity was
extended by gauging the Maxwell algebra as well as the so called AdS-Maxwell
algebra. \ In this letter it is shown that the actions for these
four-dimensional extended Einstein gravities can be obtained from the
five-dimensional Chern-Simons gravities actions by using the Randall-Sundrum
compactification procedure. It is found that the In\"{o}n\"{u}-Wigner
contraction procedure, in the Weimar-Woods sense, can be used both to obtain
the Maxwell-Chern-Simons action from the $AdS\mathcal{-}Maxwell$%
-Chern-Simons action and to obtain the Maxwell extension of Einstein gravity
in $4D$ from the four-dimensional extended $AdS$-$Maxwell$-Einstein-Hilbert
action. It is also shown that the extended four-dimensional gravities
belongs to the Horndeski family of scalar-tensor theories.
\end{abstract}

\section{\textbf{Introduction}}

The problem of adding a cosmological term, including abelian gauge fields,
to Einstein's field equations was treated in\textbf{\ }Refs. \cite%
{azcarr,azcarr1} by gauging the $D=4$ Maxwell algebra \cite{maxw1,maxw2},
which can also be obtained from (A)dS algebra using the Lie algebra
expansion procedure developed in Refs. \cite{exp1,exp2,exp3,exp4}. \ This
method allows also to derive the so called $AdS$-Maxwell algebra \cite%
{sor1,sor3,ss}, whose generators satisfy the following commutation relations%
\footnote{%
This algebra was also reobtained in Ref. \cite{gomis} from Maxwell algebra
through a procedure known as deformation.},

\begin{align}
\left[ J_{ab},J_{cd}\right] & =\eta _{bc}J_{ad}+\eta _{ad}J_{bc}-\eta
_{ac}J_{bd}-\eta _{bd}J_{ac},  \notag \\
\left[ J_{ab},Z_{cd}\right] & =\eta _{bc}Z_{ad}+\eta _{ad}Z_{bc}-\eta
_{ac}Z_{bd}-\eta _{bd}Z_{ac},  \notag \\
\left[ Z_{ab},Z_{cd}\right] & =\eta _{bc}Z_{ad}+\eta _{ad}Z_{bc}-\eta
_{ac}Z_{bd}-\eta _{bd}Z_{ac},  \notag \\
\left[ J_{ab},P_{c}\right] & =\eta _{bc}P_{a}-\eta _{ac}P_{b},\text{ \ \ \ }%
\left[ P_{a},P_{b}\right] =Z_{ab},  \notag \\
\left[ Z_{ab},P_{c}\right] & =\eta _{bc}P_{a}-\eta _{ac}P_{b}.  \label{eje3}
\end{align}

This algebra was found in \cite{sor1,sor3}, where was called "semisimple
extended Poincar\'{e} algebra". In Ref. \cite{gomis} this algebra was
obtained from Maxwell algebra through of the deformation procedure and
called "AdS-Maxwell algebra" and in Ref. \cite{ss} it was obtained from the
expansion procedure and called "AdS-Lorentz $(AdS\mathcal{L}_{N})$ algebra".

The Chern-Simons gravity has been extensively investigated within several
theoretical frameworks. In three-dimensional spacetime, the Chern-Simons
gravity invariant under (A)dS algebra is equivalent to the Einstein-Hilbert
action with a cosmological constant \cite{witten}. Furthermore, in the
context of higher dimensions, the (A)dS-Chern-Simons gravity can be obtained
by properly selecting the coefficients in the Lovelock theory \cite{lov,tron}%
. These results have also been generalized for symmetries that are given by
expansions and contractions of the (A)dS algebra \cite{isa,edel,con,con1}.
Nevertheless, the formulation of Chern-Simons gravity is limited to odd
dimensions. On the other hand, it was shown in Ref.\cite{char} that several
terms of (D=4) Horndeski action \cite{horn} emerged from the Kaluza-Klein
dimensional reduction of Lovelock theory. Consequently, it is interesting to
investigate the effective theories derived from dimensional reductions of
extended Chern-Simons gravity, which incorporate non-abelian fields \cite%
{ban,gon,mor,car,salg}.

From the commutation relations (\ref{eje3}) we can note that the set $%
\mathfrak{I}=\left( P_{a},Z_{ab}\right) $ satisfies the conditions $\left[ 
\mathfrak{I},\mathfrak{I}\right] \subset \mathfrak{I}$, $\left[ so(3,1),%
\mathfrak{I}\right] \subset \mathfrak{I}$, i.e. $\mathfrak{I}$ is an ideal
of the $AdS$-Maxwell algebra, which means that the $AdS$-Maxwell$%
=so(3,1)\uplus \mathfrak{I}$.

The main purpose of this article is to show that the four-dimensional
extended Einstein gravity with a cosmological term including non-abelian
gauge fields,\textbf{\ }found in \cite{car,salg}, may derive from
five-dimensional $AdS$-Maxwell-Chern-Simons gravity. This can be achieved by
replacing a Randall-Sundrum type metric in the five-dimensional Chern-Simons
action for $AdS$-Maxwell algebra. The same procedure is used to obtain the
four-dimensional extended Einstein gravity, with a cosmological term
including Abelian gauge fields, found in Refs. \cite{azcarr,azcarr1}, from
the five-dimensional Maxwell-Chern-Simons gravity action.

This paper is organized as follows: In Section $2$ we consider a brief
review of the construction of the $AdS$-Maxwell-Chern-Simons Lagrangian
gravity and then we obtain the Maxwell-Chern-Simons Lagrangian using the In%
\"{o}n\"{u}-Wigner contraction procedure in the Weimar-Woods sense. In
Section $3$ we apply the so called Randall-Sundrum compactification
procedure to the $AdS$-Maxwell-Chern-Simons Lagrangian gravity to obtain the
extended four-dimensional Einstein-Hilbert action with a cosmological term
including non-abelian gauge fields. This procedure is also applied to the
Maxwell-Chern-Simons gravity action to obtain an action for the extended \
four-dimensional Einstein gravity, which coincides, except for some
coefficients, with the action obtained some years ago in references \cite%
{azcarr,azcarr1}. Finally, it is shown in Section $4$ that the four
dimensional actions obtained in Section $3$ belong to the family of
Horndeski actions. Three appendices and concluding remarks end this article.

\section{\textbf{Maxwell- Chern-Simons action from AdS-Maxwell-Chern-Simons
gravity}}

In this section we use the dual $S$-expansion procedure \cite{exp3} to find
the five-dimensional Chern--Simons Lagrangian invariant under the $AdS$%
-Maxwell algebra \cite{ss}, and then using the In\"{o}n\"{u}-Wigner
contraction procedure we find the Chern-Simons Lagrangian for the Maxwell
algebra.

\subsection{$AdS$\textbf{-Maxwell}-\textbf{Chern-Simons gravity action}}

In order to write down an $AdS$\textbf{-}Maxwell-Chern--Simons Lagrangian,
we start from the $AdS$-Maxwell algebra valued one-form gauge connection%
\begin{equation*}
A=\frac{1}{l}e^{a}P_{a}+\frac{1}{2}\omega ^{ab}J_{ab}+\frac{1}{2}%
k^{ab}Z_{ab},
\end{equation*}%
where $a,b=0,1,2,3,4$ are tangent space indices raised and lowered with the
Minkowski metric $\eta _{ab}$, and where 
\begin{equation*}
e^{a}=e_{\mu }^{a}dx^{\mu },\text{ \ }\omega ^{ab}=\omega _{\mu
}^{ab}dx^{\mu },\text{ \ }k^{ab}=k_{\mu }^{ab}dx^{\mu },
\end{equation*}%
are the $e_{\mu }^{a}$ f\'{u}nfbein, the $\omega _{\mu }^{ab}$ spin
connection and the $k_{\mu }^{ab}$ new non-abelian gauge fields. The
corresponding associated curvature $2$-form, is given by 
\begin{equation}
F=\frac{1}{l}\mathcal{T}^{a}P_{a}+\frac{1}{2}R^{ab}J_{ab}+\frac{1}{2}%
F^{ab}Z_{ab},  \label{12}
\end{equation}%
with 
\begin{align}
\mathcal{T}^{a}& =T^{a}+k_{\;c}^{a}e^{c},  \notag \\
R^{ab}& =d\omega ^{ab}+\omega _{\,c}^{a}\omega ^{cb},  \notag \\
F^{ab}& =Dk^{ab}+k_{\;\;c}^{[a}k^{c|b]}+\frac{1}{l^{2}}e^{a}e^{b}.
\label{13}
\end{align}

In this point, it might be of interest to remember that: $\left( i\right) $\
clearly $l$\ could be eliminated by\textbf{\ }absorbing it in the definition
of the vielbein, but then the space-time metric $g_{\mu \nu }$\ would no
longer be related to $e^{a}$\ through the relation $g_{\mu \nu }=\eta
_{ab}e_{\mu }^{\text{ }a}e_{\nu }^{\text{ }b}$; $\left( ii\right) $\ the
interpretation of the $l$\ parameter as a parameter related to the radius of
curvature of the $AdS$\ space-time, is inherited for the space-time whose
symmetries are described by the Maxwell algebra.

On the another hand, it could also be interesting to observe that $J_{ab}$\
are still Lorentz generators, but $P_{a}$\ are no longer $AdS$\ boosts. In
fact, $\left[ P_{a},P_{b}\right] =Z_{ab}$. However $e^{a}$\ still transforms
as a vector under Lorentz transformations, as it must, in order to recover
gravity in this scheme.

A Chern-Simons Lagrangian in $D=5$\ dimensions is defined to be the
following local function of a one-form gauge connection $A$: 
\begin{equation}
\mathcal{L}_{ChS}^{\left( 5D\right) }\left( A\right) =\left\langle AF^{2}-%
\frac{1}{2}A^{3}F+\frac{1}{10}A^{5}\right\rangle ,  \label{lcs}
\end{equation}%
where $\left\langle \cdots \right\rangle $\ denotes an invariant tensor for
the corresponding Lie algebra$,$\ $F=dA+AA$\ is the corresponding two-form
curvature \cite{zan}.

Using Theorem VII.2 of Ref. \cite{exp1}, it is possible to show that the
only non-vanishing components of an invariant tensor for the $AdS$-Maxwell
algebra are given by%
\begin{align*}
\left\langle J_{ab}J_{cd}P_{e}\right\rangle & =\frac{4}{3}\alpha
_{1}l^{3}\varepsilon _{abcde}, \\
\left\langle Z_{ab}Z_{cd}P_{e}\right\rangle & =\frac{4}{3}\alpha
_{1}l^{3}\varepsilon _{abcde}, \\
\left\langle J_{ab}Z_{cd}P_{e}\right\rangle & =\frac{4}{3}\alpha
_{1}l^{3}\varepsilon _{abcde},
\end{align*}%
where $\alpha _{1}$ is an arbitrary constant of dimensions $\left[ \text{%
length}\right] ^{-3}$.

Using the dual $S$-expansion procedure in terms of Maurer--Cartan forms \cite%
{exp3}, we find that the five-dimensional Chern--Simons Lagrangian invariant
under the $AdS$-Maxwell algebra is given by \cite{ss}%
\begin{align}
\mathcal{L}_{\mathrm{ChS}}^{\left( \mathrm{AdS}\mathcal{M}\right) }& =\alpha
_{1}\varepsilon _{abcde}\left\{ l^{2}R^{ab}R^{cd}e^{e}+l^{2}\left(
Dk^{ab}\right) \left( Dk^{cd}\right) e^{e}\right.  \notag \\
& +l^{2}k_{\text{ \ }f}^{a}k^{fb}k_{\text{ \ }g}^{c}k^{gd}e^{e}+\frac{1}{%
5l^{2}}e^{a}e^{b}e^{c}e^{d}e^{e}+2l^{2}R^{ab}k_{\text{ \ }f}^{c}k^{fd}e^{e} 
\notag \\
& +\frac{2}{3}\left( Dk^{ab}\right) e^{c}e^{d}e^{e}+2l^{2}R^{ab}\left(
D_{\omega }k^{cd}\right) e^{e}  \notag \\
& \left. +2l^{2}\left( Dk^{ab}\right) k_{\text{ \ }f}^{c}k^{fd}e^{e}+\frac{2%
}{3}R^{ab}e^{c}e^{d}e^{e}+\frac{2}{3}k_{\text{ \ }%
f}^{a}k^{fb}e^{c}e^{d}e^{e}\right\} ,  \label{1}
\end{align}%
where $\alpha _{1}$ is a parameter of the theory, $R^{ab}=\text{d}\omega
^{ab}+\omega _{\text{ \ }c}^{a}\omega ^{cb}$ correspond to the curvature $2$%
-form in the first-order formalism related to the spin connection $1$-form, $%
e^{a}$ is the vielbein $1$-form, and $k^{ab}$ $1$-form are others gauge
fields presents in the theory.

\subsection{\textbf{Maxwell Chern-Simons action}}

Keeping in mind that the Maxwell algebra can be obtained from $AdS$-Maxwell
algebra by a generalized In\"{o}n\"{u}--Wigner contraction \cite{ss,luk},
the natural question is how to obtain the corresponding Lagrangian for the
Maxwell algebra from the Lagrangian for the $AdS$-Maxwell algebra?. We find
that it is also possible to obtain this relation using the same procedure
which was applied to the algebras. In fact, carrying out the rescaling of
the generators $P_{a}\rightarrow \xi P_{a}$, $Z_{ab}\rightarrow \xi
^{2}Z_{ab}$ and of the fields $e^{a}\rightarrow \xi ^{-1}e^{a}$, $%
k^{ab}\rightarrow \xi ^{-2}k^{ab}$ in the Lagrangian (\ref{1}) we obtain

\begin{align}
L_{\mathrm{ChS}}^{\left( \mathcal{M}\right) }& =\alpha _{1}l^{2}\varepsilon
_{abcde}R^{ab}R^{cd}e^{e}+\frac{2}{3}\alpha _{1}\varepsilon
_{abcde}R^{ab}e^{c}e^{d}e^{e}  \notag \\
& +\frac{\alpha _{1}}{5l^{2}}\varepsilon _{abcde}e^{a}e^{b}e^{c}e^{d}e^{e}+%
\frac{2}{3}\alpha _{1}\varepsilon _{abcde}\left( D_{\omega }k^{ab}\right)
e^{c}e^{d}e^{e}  \notag \\
& +\alpha _{1}l^{2}\varepsilon _{abcde}\left( D_{\omega }k^{ab}\right)
\left( D_{\omega }k^{cd}\right) e^{e}+2\alpha _{1}\varepsilon
_{abcde}R^{ab}\left( D_{\omega }k^{cd}\right) e^{e},  \label{2'}
\end{align}%
which corresponds to the five-dimensional Chern-Simons Lagrangian for the
Maxwell algebra.

\section{\textbf{Extended four-dimensional Einstein-Hilbert action from }$%
AdS $\textbf{-Maxwell-Chern-Simons gravity action}}

In order to obtain an action for a $4$-dimensional gravity theory from the
Chern-Simons action for $AdS$-Maxwell \'{a}lgebra we will consider the
following $5$-dimensional Randall Sundrum type metric \cite%
{randall,randall1,rd,gomez} 
\begin{eqnarray}
ds^{2} &=&e^{2f(\phi )}\tilde{g}_{\mu \nu }(\tilde{x})d\tilde{x}^{\mu }d%
\tilde{x}^{\nu }+r_{c}^{2}d\phi ^{2}  \notag \\
&=&e^{2f(\phi )}\tilde{\eta}_{mn}\tilde{e}^{m}\tilde{e}^{n}+r_{c}^{2}d\phi
^{2},
\end{eqnarray}%
where $e^{2f(\phi )}$ is the so-called "warp factor", and $r_{c}$ is the
so-called "compactification radius" of the extra dimension, which is
associated with the coordinate $0\leqslant \phi <2\pi $. The symbol $\sim $
denotes $4$-dimensional quantities related to the space-time $\Sigma _{4}.$
We will use the usual notation, 
\begin{eqnarray}
x^{\alpha } &=&\left( \tilde{x}^{\mu },\phi \right) ;\text{ \ \ \ \ }\alpha
,\beta =0,...,4;\text{ \ \ \ \ }a,b=0,...,4;  \notag \\
\mu ,\nu &=&0,...,3;\text{ \ \ \ \ }m,n=0,...,3;  \notag \\
\eta _{ab} &=&diag(-1,1,1,1,1);\text{ \ \ \ \ }\tilde{\eta}%
_{mn}=diag(-1,1,1,1).
\end{eqnarray}

This allows us, for example, to write 
\begin{eqnarray}
e^{m}(\phi ,\tilde{x}) &=&e^{f(\phi )}\tilde{e}^{m}(\tilde{x})=e^{f(\phi )}%
\tilde{e}_{\text{ }\mu }^{m}(\tilde{x})d\tilde{x}^{\mu };\text{ \ \ }%
e^{4}(\phi )=r_{c}d\phi .  \notag \\
k^{mn}(\phi ,\tilde{x}) &=&\tilde{k}^{mn}(\tilde{x})\text{, }k^{m4}=k^{4m}=0%
\text{\ ,}  \label{s2}
\end{eqnarray}%
where, following Randall and Sundrum \cite{randall,randall1} matter fields
are null in the fifth dimension.

From the vanishing torsion condition%
\begin{equation}
T^{a}=de^{a}+\omega _{\text{ }b}^{a}e^{b}=0,  \label{2t}
\end{equation}%
we obtain 
\begin{equation}
\omega _{\text{ }b\alpha }^{a}=-e_{\text{ }b}^{\beta }\left( \partial
_{\alpha }e_{\text{ }\beta }^{a}-\Gamma _{\text{ }\alpha \beta }^{\gamma }e_{%
\text{ }\gamma }^{a}\right) ,  \label{3t}
\end{equation}%
where $\Gamma _{\text{ }\alpha \beta }^{\gamma }$ is the Christoffel symbol.

From equations (\ref{s2}) and (\ref{2t}), we find%
\begin{equation}
\omega _{\text{ }4}^{m}=\frac{e^{f}f^{\prime }}{r_{c}}\tilde{e}^{m},\text{
with }f^{\prime }=\frac{\partial f}{\partial \phi },  \label{102t}
\end{equation}%
and the $4$-dimensional vanishing torsion condition 
\begin{equation}
\tilde{T}^{m}=\tilde{d}\tilde{e}^{m}+\tilde{\omega}_{\text{ }n}^{m}\tilde{e}%
^{n}=0,\text{ with \ }\tilde{\omega}_{\text{ }n}^{m}=\omega _{\text{ }n}^{m}%
\text{ \ and }\tilde{d}=d\tilde{x}^{\mu }\frac{\partial }{\partial \tilde{x}%
^{\mu }}.  \label{1030t}
\end{equation}

From (\ref{102t}), (\ref{1030t}) and the Cartan's second structural
equation, $R^{ab}=d\omega ^{ab}+\omega _{\text{ }c}^{a}\omega ^{cb}$, we
obtain the components of the $2$-form curvature%
\begin{equation}
R^{m4}=\frac{e^{f}}{r_{c}}\left( f^{\prime 2}-f^{\prime \prime }\right)
d\phi \tilde{e}^{m},\text{ \ }R^{mn}=\tilde{R}^{mn}-\left( \frac{%
e^{f}f^{\prime }}{r_{c}}\right) ^{2}\tilde{e}^{m}\tilde{e}^{n},\text{\ }
\label{105t}
\end{equation}%
where the $4$-dimensional $2$-form curvature is given by%
\begin{equation}
\tilde{R}^{mn}=\tilde{d}\tilde{\omega}^{mn}+\tilde{\omega}_{\text{ }p}^{m}%
\tilde{\omega}^{pn}.
\end{equation}

From equation (\ref{1}) we can see that the Lagrangian contains ten terms
that we will denote as $L_{1},L_{2},\cdots ,L_{10},$ where $L_{1}$
corresponds to the Gauss-Bonnet term, $L_{4}$ corresponds to the
cosmological term, $L_{9}$ correspond to the Einstein-Hilbert term. In fact,
following Ref. \cite{rd,gomez} we replace (\ref{s2}) and (\ref{105t}) in (%
\ref{1}), and using $\tilde{\varepsilon}_{mnpq}=\varepsilon _{mnpq4}$, we
obtain%
\begin{eqnarray}
&&L_{1}=\alpha _{1}l^{2}\varepsilon _{abcde}R^{ab}R^{cd}e^{e}  \notag \\
&=&\alpha _{1}l^{2}r_{c}d\phi \left\{ \tilde{\varepsilon}_{mnpq}\tilde{R}%
^{mn}\tilde{R}^{pq}-\left( \frac{2e^{2f}}{r_{c}^{2}}\right) \left(
3f^{\prime 2}+2f^{\prime \prime }\right) \tilde{\varepsilon}_{mnpq}\tilde{R}%
^{mn}\tilde{e}^{p}\tilde{e}^{q}\right.  \notag \\
&&\text{ \ \ \ \ \ \ \ \ \ \ \ \ \ \ }\left. +\left( \frac{e^{4f}}{r_{c}^{4}}%
f^{\prime 2}\right) \left( 5f^{\prime 2}+4f^{\prime \prime }\right) \tilde{%
\varepsilon}_{mnpq}\tilde{e}^{m}\tilde{e}^{n}\tilde{e}^{p}\tilde{e}%
^{q}\right\} ,  \label{l1}
\end{eqnarray}%
\begin{equation}
L_{2}=\alpha _{1}l^{2}r_{c}d\phi \tilde{\varepsilon}_{mnpq}Dk^{mn}Dk^{pq},
\label{l2}
\end{equation}%
\begin{equation}
L_{3}=\alpha _{1}\ell ^{2}r_{c}d\phi \tilde{\epsilon}_{mnpq}\tilde{k}_{\
f}^{m}\tilde{k}^{fn}\tilde{k}_{\ g}^{p}\tilde{k}^{gq},  \label{l3}
\end{equation}%
\begin{equation}
L_{4}=\frac{\alpha _{1}}{l^{2}}r_{c}d\phi e^{4f}\tilde{\varepsilon}_{mnpq}%
\tilde{e}^{m}\tilde{e}^{n}\tilde{e}^{p}\tilde{e}^{q},  \label{l4}
\end{equation}

\begin{eqnarray}
L_{5} &=&2\alpha _{1}\ell ^{2}r_{c}d\phi \left[ \tilde{\epsilon}_{mnpq}%
\tilde{R}^{mn}\tilde{k}_{\ f}^{p}\tilde{k}^{fq}-\frac{e^{2f(\phi )}}{%
r_{c}^{2}}\left( 2f^{\prime \prime }+3f^{\prime 2}\right) \tilde{\epsilon}%
_{mnpq}\tilde{k}_{\ f}^{m}\tilde{k}^{fn}\tilde{e}^{p}\tilde{e}^{q}\right] , 
\notag \\
&&  \label{l5}
\end{eqnarray}

\begin{equation}
L_{6}=2\alpha _{1}r_{c}d\phi e^{2f(\phi )}\tilde{\epsilon}_{mnpq}\left(
D_{\omega }\tilde{k}^{mn}\right) \tilde{e}^{p}\tilde{e}^{q},  \label{l6}
\end{equation}%
\begin{equation}
L_{7}=2\alpha _{1}l^{2}r_{c}d\phi \tilde{\varepsilon}_{mnpq}\left\{ \tilde{R}%
^{mn}Dk^{pq}-\frac{e^{2f}}{r_{c}^{2}}\left( 3f^{\prime 2}+2f^{\prime \prime
}\right) Dk^{mn}\tilde{e}^{p}\tilde{e}^{q}\right\} ,  \label{l7}
\end{equation}

\begin{equation}
L_{8}=2\alpha _{1}\ell ^{2}r_{c}d\phi \tilde{\epsilon}_{mnpq}\left(
D_{\omega }\tilde{k}^{mn}\right) \tilde{k}_{\ f}^{p}\tilde{k}^{fq},
\label{l8}
\end{equation}

\begin{eqnarray}
L_{9} &=&\frac{2}{3}\alpha _{1}r_{c}d\phi \left\{ \left( 3e^{2f}\right) 
\tilde{\varepsilon}_{mnpq}\tilde{R}^{mn}\tilde{e}^{p}\tilde{e}^{q}\right. 
\notag \\
&&-\left. \left( \frac{e^{4f}}{r_{c}^{2}}\right) \left( 5f^{\prime
2}+2f^{\prime \prime }\right) \tilde{\varepsilon}_{mnpq}\tilde{e}^{m}\tilde{e%
}^{n}\tilde{e}^{p}\tilde{e}^{q}\right\} ,  \label{l9}
\end{eqnarray}

\begin{equation}
L_{10}=2\alpha _{1}r_{c}d\phi e^{2f(\phi )}\tilde{\varepsilon}_{mnpq}\tilde{k%
}_{\ f}^{m}\tilde{k}^{fn}\tilde{e}^{p}\tilde{e}^{q}.  \label{l10}
\end{equation}

By replacing (\ref{l1}-\ref{l10}) in (\ref{1}) and integrating over the
fifth dimension we find 
\begin{eqnarray}
S_{4D}^{AdS\mathcal{M}} &=&\int_{\Sigma _{4}}A\tilde{\varepsilon}_{mnpq}%
\left[ \tilde{R}^{mn}\tilde{e}^{p}\tilde{e}^{q}+\tilde{k}_{\ f}^{m}\tilde{k}%
^{fn}\tilde{e}^{p}\tilde{e}^{q}+D\tilde{k}^{mn}\tilde{e}^{p}\tilde{e}^{q}%
\right]  \notag \\
&&+B\tilde{\varepsilon}_{mnpq}\tilde{e}^{m}\tilde{e}^{n}\tilde{e}^{p}\tilde{e%
}^{q}+C\tilde{\varepsilon}_{mnpq}\left[ D\tilde{k}^{mn}D\tilde{k}^{pq}+%
\tilde{k}_{\ f}^{m}\tilde{k}^{fn}\tilde{k}_{\ g}^{p}\tilde{k}^{gq}\right. 
\notag \\
&&\left. +2\tilde{R}^{mn}\tilde{k}_{\ f}^{p}\tilde{k}^{fq}+2D\tilde{k}^{mn}%
\tilde{k}_{\ f}^{p}\tilde{k}^{fq}\right] +sourface\text{ }terms,  \label{l11}
\end{eqnarray}

where,%
\begin{eqnarray}
A &=&2\alpha _{1}r_{c}\int_{0}^{2\pi }e^{2f(\phi )}\left[ 1-\frac{\ell ^{2}}{%
r_{c}^{2}}\left( 2f^{\prime \prime }+3f^{\prime 2}\right) \right] d\phi ,
\label{l12} \\
&=&\frac{2\pi \alpha _{1}\left( \ell ^{2}+r_{c}^{2}\right) }{r_{c}}
\label{l14} \\
B &=&\alpha _{1}r_{c}\int_{0}^{2\pi }e^{4f(\phi )}\left[ \frac{\ell ^{2}}{%
r_{c}^{4}}\left( 4f^{\prime \prime }+5f^{\prime 2}\right) f^{\prime 2}+\frac{%
1}{\ell ^{2}}-\frac{2}{3r_{c}^{2}}\left( 2f^{\prime \prime }+5f^{\prime
2}\right) \right] d\phi  \notag \\
&=&\frac{\pi \alpha _{1}}{4\ell ^{2}r_{c}^{3}}\left[ 3r_{c}^{4}+2\ell
^{2}r_{c}^{2}-\ell ^{4}\right] ,  \label{l13} \\
C &=&2\pi \alpha _{1}\ell ^{2}r_{c}
\end{eqnarray}%
with $f(\phi )$ an arbitrary and continuously differentiable function. Since
we are working with a cylindrical variety, we have chosen (non-unique
choice) $f(\phi )=\ln \left( \sin \phi \right) .$

From the action (\ref{l11}) we see that it includes\textbf{\ }non-Abelian
fields $\tilde{k}_{\mu }^{mn}$, which could be interpreted as non-Abelian
gauge field that driven inflation (see e.g. \cite{mot3,mot4,mot5,mot6,mot7}).

\subsection{\textbf{Maxwell Einstein gravity from Maxwell Chern-Simons action%
}}

In order to obtain an action for a $4$-dimensional gravity theory from the
Chern-Simons action for Maxwell \'{a}lgebra we follow the same procedure
used in the previous section. In fact, replacing (\ref{s2}) and (\ref{105t})
in (\ref{2'}), and using $\tilde{\varepsilon}_{mnpq}=\varepsilon _{mnpq4}$,

\begin{eqnarray}
S_{\mathcal{M}}^{GEH} &=&\int_{\Sigma _{4}}A\varepsilon _{mnpq}\left[ \tilde{%
R}^{mn}\tilde{e}^{p}\tilde{e}^{q}+D\tilde{k}^{mn}\tilde{e}^{p}\tilde{e}^{q}%
\right]  \notag \\
&&+B\varepsilon _{mnpq}\tilde{e}^{m}\tilde{e}^{n}\tilde{e}^{p}\tilde{e}%
^{q}+2\pi \alpha _{1}\ell ^{2}r_{c}\varepsilon _{mnpq}D\tilde{k}^{mn}D\tilde{%
k}^{pq},  \notag \\
&&  \label{l16}
\end{eqnarray}%
where the coeficients $A$ and $B$ are given by eqs. (\ref{l14}, \ref{l13}).
This action matches with the action (\ref{azc}), which correspond to the
equation $\left( 29\right) $ of reference \cite{azcarr}, as long as $%
A=-1/2\kappa =\lambda /2\kappa \Lambda $, $B=\lambda /4\kappa $, $2\pi
\alpha _{1}\ell ^{2}r_{c}=\lambda /2\kappa \Lambda ^{2}.$ This means that
the action (\ref{l16}) and the action (\ref{azc}) coincide only if $\Lambda
=1/l^{2}$, that is if $\lambda =-1/l^{2}$.

It is of interest to note that the action (\ref{l16}) can be obtained from
the action (\ref{l11}) using the generalized In\"{o}n\"{u}--Wigner
contraction, namely, carrying out the rescaling of the generators $%
P_{a}\rightarrow \xi P_{a}$, $Z_{ab}\rightarrow \xi ^{2}Z_{ab}$ and of the
fields $e^{a}\rightarrow \xi ^{-1}e^{a}$, $k^{ab}\rightarrow \xi ^{-2}k^{ab}$
in the Lagrangian (\ref{l16}).

\section{\textbf{From} \textbf{AdS-Maxwell gravity to scalar-tensor theory}}

In this Section it is found that the four-dimensional actions obtained from
Chern-Simons gravity actions invariants under the so called generalized
(A)dS-Maxwell symmetries belongs to a larges class of theories known as
Horndeski theories (see Appendix 2).

The non-abelian gauge field in four-dimensional spacetime is a rank-three
tensor with two anti-symmetric indices, $\tilde{k}_{[mn]p}$. This means that
it has $24$ degrees of freedom (d.o.f.). We can decompose this field with
respect to the Lorentz group into three irreducible tensors \cite{mc},\cite%
{cap}, namely 
\begin{equation}
\tilde{k}_{[mn]p}=-\frac{1}{3}\left( \tilde{k}_{m}\eta _{np}-\tilde{k}%
_{n}\eta _{mp}\right) -\frac{1}{6}\varepsilon _{mnpq}S^{q}+q_{mnp},
\label{h}
\end{equation}%
where, the trace vector $\tilde{k}_{m}\equiv \tilde{k}_{\text{ }mn}^{n}$ has 
$4$ d.o.f.. The axial vector $S^{q}$ possesses $4$ d.o.f., while $q_{mnp}$
exhibits $16$ d.o.f., representing the traceless and non-totally
anti-symmetric component of the tensor. In order to obtain an effective
scalar-tensor theory from AdS-Maxwell gravity, it is necessary to consider a
single additional degree of freedom; accordingly, we set a value of zero to
both the axial vector and the $q$ tensor. Furthermore, we choose the trace
vector as $\tilde{k}_{m}=\tilde{e}_{m}^{\text{ \ }\mu }D_{\mu }\varphi $
that is, we postulate that $\tilde{k}_{m}$ is the gradient of a scalar
field, therefore depending completely on one degree of freedom. Thus, we
have that

\begin{equation}
\tilde{k}^{mn}\equiv \tilde{k}_{\text{ \ \ \ }p}^{mn}\tilde{e}^{p}=-\frac{1}{%
3}\left( \tilde{k}^{m}\tilde{e}^{n}-\tilde{k}^{n}\tilde{e}^{m}\right) .\text{
\ }  \label{h0'}
\end{equation}%
Taking this ansatz into account we have that the eight terms of (\ref{l11})
take the form (see Appendix 3) 
\begin{equation}
A\tilde{\epsilon}_{mnpq}\tilde{R}^{mn}\tilde{e}^{p}\tilde{e}^{q}=4A\sqrt{-%
\tilde{g}}\tilde{R}d^{4}\tilde{x}.  \label{h1}
\end{equation}%
\begin{equation}
A\tilde{\epsilon}_{mnpq}\tilde{k}_{\ f}^{m}\tilde{k}^{fn}\tilde{e}^{p}\tilde{%
e}^{q}=-\frac{4}{3}AD_{\mu }\varphi D^{\mu }\varphi \sqrt{\tilde{g}}d^{4}x.
\label{h6}
\end{equation}

\begin{equation}
A\tilde{\epsilon}_{mnpq}D\tilde{k}^{mn}\tilde{e}^{p}\tilde{e}^{q}=-4AD_{\mu
}D^{\mu }\varphi \sqrt{-\tilde{g}}d^{4}\tilde{x}  \label{h8}
\end{equation}

\begin{equation}
B\tilde{\epsilon}_{mnpq}\tilde{e}^{m}\tilde{e}^{n}\tilde{e}^{p}\tilde{e}%
^{q}=24B\sqrt{-\tilde{g}}d^{4}\tilde{x}.  \label{h9}
\end{equation}

\begin{equation}
C\tilde{\varepsilon}_{mnpq}D\tilde{k}^{mn}D\tilde{k}^{pq}=\frac{8C}{9}%
\left\{ \left( D_{\alpha }D^{\alpha }\varphi \right) ^{2}-D_{\gamma }D^{\nu
}\varphi D_{\nu }D^{\gamma }\varphi \right\} \sqrt{-\tilde{g}}d^{4}\tilde{x}
\label{h10'}
\end{equation}%
\begin{equation}
C\tilde{\varepsilon}_{mnpq}\tilde{k}_{\ f}^{m}\tilde{k}^{fn}\tilde{k}_{\
g}^{p}\tilde{k}^{gq}=0  \label{h15}
\end{equation}

\begin{equation}
2C\tilde{\varepsilon}_{mnpq}\tilde{R}^{mn}\tilde{k}_{\ f}^{p}\tilde{k}^{fq}=%
\frac{-8C}{3^{2}}\sqrt{-\tilde{g}}d^{4}\tilde{x}D^{\alpha }\varphi D^{\beta
}\varphi \tilde{G}_{\alpha \beta }-\frac{8C}{3^{2}}\sqrt{-\tilde{g}}d^{4}%
\tilde{x}D_{\nu }\varphi D^{\nu }\varphi \tilde{R}  \label{h21}
\end{equation}

\begin{equation}
2C\tilde{\varepsilon}_{mnpq}D\tilde{k}^{mn}\tilde{k}_{\ f}^{p}\tilde{k}^{fq}=%
\sqrt{-\tilde{g}}d^{4}\tilde{x}\frac{8C}{3^{3}}D_{\gamma }A^{\gamma },\text{
con \ \ }A^{\gamma }=D_{\alpha }\varphi D^{\alpha }\varphi D^{\gamma
}\varphi .  \label{h26}
\end{equation}

Replacing (\ref{h1}, \ref{h6}, \ref{h8}, \ref{h9},\ref{h10'}, \ref{h15}, \ref%
{h21}, \ref{h26}) en (\ref{l11}), we find:

\begin{eqnarray}
S_{4D}^{AdS\mathcal{M}} &=&\int_{\Sigma _{4}}d^{4}\tilde{x}\sqrt{-\tilde{g}}%
\left\{ \left( 4A-\frac{8C}{9}D_{\nu }\varphi D^{\nu }\varphi \right) \tilde{%
R}-\frac{4}{3}AD_{\mu }\varphi D^{\mu }\varphi -4AD_{\mu }D^{\mu }\varphi
\right.  \notag \\
&&+24B+\frac{8C}{9}\left[ \left( D_{\alpha }D^{\alpha }\varphi \right)
^{2}-D_{\gamma }D_{\nu }\varphi D^{\nu }D^{\gamma }\varphi \right]  \notag \\
&&\left. +\frac{8C}{9}\phi D^{\alpha }D^{\beta }\varphi \tilde{G}_{\alpha
\beta }+surface\text{ }terms\right\} ,  \label{h29}
\end{eqnarray}%
where, we have used the followin identities 
\begin{eqnarray}
D^{\alpha }\varphi D^{\beta }\varphi \tilde{G}_{\alpha \beta } &=&D^{\alpha }%
\left[ \varphi D^{\beta }\varphi \tilde{G}_{\alpha \beta }\right] -\varphi
D^{\alpha }D^{\beta }\varphi \tilde{G}_{\alpha \beta },  \notag \\
D^{\alpha }\tilde{G}_{\alpha \beta } &=&0.
\end{eqnarray}

Comparing these results with the equations (\ref{a3-1}) of Appendix $2$, we
can see that the Lagrangian (\ref{h29}) can be written as

\begin{eqnarray}
S_{4D}^{AdS\mathcal{M}} &=&\int_{\Sigma _{4}}d^{4}\tilde{x}\sqrt{-\tilde{g}}%
\left\{ G_{2}\left( \varphi ,X\right) +G_{3}\left( \varphi ,X\right) D_{\mu
}D^{\mu }\varphi +G_{4}\left( \varphi ,X\right) \tilde{R}\right.  \notag \\
&&-2G_{4X}\left( \varphi ,X\right) \left[ \left( D_{\alpha }D^{\alpha
}\varphi \right) ^{2}-\left( D^{\mu }D^{\nu }\varphi \right) \left( D_{\mu
}D_{_{\nu }}\varphi \right) \right]  \notag \\
&&\left. +G_{5}\left( \varphi ,X\right) \tilde{G}_{\mu \upsilon }D^{\mu
}D^{\nu }\varphi +surface\text{ }terms\right\} ,  \label{h27}
\end{eqnarray}%
where%
\begin{eqnarray}
G_{2}\left( \varphi ,X\right) &=&-\frac{4}{3}AD_{\mu }\varphi D^{\mu
}\varphi +24B  \notag \\
G_{3}\left( \varphi ,X\right) &=&-4A  \notag \\
G_{4}\left( \varphi ,X\right) &=&4A-\frac{8C}{9}D_{\nu }\varphi D^{\nu
}\varphi  \notag \\
G_{4X}\left( \varphi ,X\right) &=&-\frac{4C}{9}  \notag \\
G_{5}\left( \varphi ,X\right) &=&\frac{8C}{9}\varphi ,  \label{h28}
\end{eqnarray}%
This action explicitly includes Einstein-Hilbert gravity with a cosmological
constant, along with various additional components of Horndeski theory;
therefore, this is a second-order scalar-tensor theory. On the other hand,
by considering all the degrees of freedom of the non-abelian gauge field, we
will obtain an extended theory with a sector corresponding to a Horndeski
family, as well as new terms whose interpretation will depend on the
physical quantities introduced. We will investigate this theory in future
works.

It should be noted that an analogous result can be obtained in the case of
the Maxwell action (\ref{l16}), where it is straightforward to see that the
four terms of the Lagrangian corresponding to the action (\ref{l16}) are
given by ((\ref{h1}, \ref{h6}, \ref{h8}, \ref{h9}, \ref{h10'}).

\section{\textbf{Concluding Remarks }}

In this article we have obtained a four-dimensional extended Einstein
gravity with a cosmological term, including non-abelian gauge fields,\textbf{%
\ } from five-dimensional $AdS$--Maxwell-Chern-Simons gravity. This gravity
in $4D$ includes\textbf{\ }non-Abelian fields $\tilde{k}^{mn}$, which could
be interpreted as gauge fields that driven inflation. This was achieved by
making use of the Randall-Sundrum compactification procedure, which is also
used to re-obtain, from the Maxwell \ Chern-Simons gravity action, the
extended Einstein gravity in $4D$ with a cosmological term,\ which including
abelian gauge fields of Refs. \cite{azcarr,azcarr1}.

The In\"{o}n\"{u}-Wigner contraction procedure in the Weimar-Woods sense is
used both to obtain the Maxwell-Chern-Simons action from the Chern-Simon
saction for $AdS$--Maxwell algebra and to obtain the Maxwell extension of
Einstein gravity in $4D$ from $AdS$--Maxwell-Einstein-Hilbert action.

It might be of interest to note that both extensions of Einstein's gravity
with cosmological terms (which includes Abelian and non-Abelian gauge fields
respectively)\textbf{\ }are not invariant under the respective\textbf{\ }%
local transformations but only under local Lorentz transformations. Here, we
have shown that is possible to obtain this generalized four-dimensional
Einstein-Hilbert actions from the genuinely invariant five-dimensional
Chern-Simons gravities. This seems to indicate that the compactification
procedure breaks the original symmetries of the Chern-Simons actions
(Maxwell and $AdS$--Maxwell) to the Lorentz symmetry.

We have also shown that the four-dimensional actions obtained from
Chern-Simons gravity actions invariants under the so called generalized
(A)dS-Maxwell symmetries as well as under the Maxwell symmetries belongs to
the family actions for the Horndeski theory. This result allows us to
conjecture that the compactification of Chern-Simons gravities corresponding
to groups with symmetries greater than those presented here will lead to
Lagrangians that involve more Lagrangians of the basis $\mathcal{L}_{i}\left[
G_{i}\right] ,$ $i=1,2,3,4,5,\cdot \cdot \cdot \cdot \cdot $ (see Appendix $%
2 $ and Ref. \cite{gb})$.$

\section{\textbf{Appendix 1: The Maxwell algebra}}

The generators of Maxwell algebra $\left( P_{a},J_{ab},Z_{ab}\right) $
satisfying the following commutation relations \cite{maxw1,maxw2}

\begin{align}
\left[ J_{ab},J_{cd}\right] & =\eta _{bc}J_{ad}+\eta _{ad}J_{bc}-\eta
_{ac}J_{bd}-\eta _{bd}J_{ac},  \notag \\
\left[ J_{ab},P_{c}\right] & =\eta _{bc}P_{a}-\eta _{ac}P_{b},\text{ \ \ \ }%
\left[ P_{a},P_{b}\right] =Z_{ab},  \notag \\
\left[ J_{ab},Z_{cd}\right] & =\eta _{bc}Z_{ad}+\eta _{ad}Z_{bc}-\eta
_{ac}Z_{bd}-\eta _{bd}Z_{ac},  \notag \\
\left[ Z_{ab},Z_{cd}\right] & =0,\text{ \ }\left[ Z_{ab},P_{c}\right] =0.
\label{ej2}
\end{align}

From (\ref{ej2}) we see that the set $I=\left( P_{a},Z_{ab}\right) $ is an
ideal of Maxwell's algebra because $\left[ I,I\right] \subset I$, $\left[
so(3,1),I\right] \subset I$ \cite{azcarr}. This means that the Maxwell
algebra $\mathcal{M}$ is the semidirect sum of the Lorentz algebra $so(3,1)$
and the ideal $I$, that is $\mathcal{M}=so(3,1)\uplus I$ \cite{azcarr}.

The Abelan four-vector fields $k_{\mu }^{ab}$ associated with their Abelian
tensorial generators $Z_{ab}$ and the set of curvatures associated to this
gauge potentials denoted by $F_{\mu \nu }^{ab}$ allow to construct a
generalization of the Einstein-Hilbert Lagrangian given by%
\begin{eqnarray}
\mathcal{L=-} &&\frac{1}{2\kappa }\varepsilon _{abcd}R^{ab}e^{c}e^{d}+\frac{%
\lambda }{4\kappa }\varepsilon _{abcd}e^{a}e^{b}e^{c}e^{d}+\frac{\mu }{%
2\kappa }\varepsilon _{abcd}Dk^{ab}e^{c}e^{d}  \notag \\
&&+\frac{\mu ^{2}}{4\kappa \lambda }\varepsilon _{abcd}Dk^{ab}Dk^{cd},\text{
with }\mu =\lambda /\Lambda \text{ and }\Lambda =1/l^{2}  \label{azc}
\end{eqnarray}%
which corresponds to equation $\left( 29\right) $ of reference \cite{azcarr}%
, which is not invariant under local Maxwell transformations but only under
local Lorentz transformations. \ This action contain geometric Abelian gauge
fields, $k^{ab}$, playing the role of vectorial inflatons \cite{mot3}, which
contribute to a generalization of the cosmological term.

On the other hand, actions containing cosmological terms that describe
vector inflations by means of geometric non-Abelian gauge fields remain as
an open problem. The idea that dark energy could be understood by
non-abelian vector fields, that is, that non-abelian gauge fields could be
responsible for the accelerated expansion of the universe, was postulated in
references\textbf{\ }\cite{mot4,mot5,mot6,mot7}\textbf{.}

\section{\textbf{Appendix 2: Horndeski theories}}

Although Horndeski's theories have Lagrange functions that contain at most
second derivatives of a scalar field, they correspond to the more general
tenso-scalar theory that leads to second-order equations of motion. These
theories can be written as a linear combination of the following Lagrange
functions (\cite{dl},\cite{gb}) \ 
\begin{eqnarray}
\mathcal{L}_{2}\left[ G_{2}\right] &\equiv &G_{2}\left( \phi ,X\right) 
\notag \\
\mathcal{L}_{3}\left[ G_{3}\right] &\equiv &G_{3}\left( \phi ,X\right)
D_{\mu }D^{\mu }\phi  \notag \\
\mathcal{L}_{4}\left[ G_{4}\right] &\equiv &G_{4}\left( \phi ,X\right) 
\tilde{R}-2G_{4X}\left( \phi ,X\right) \left[ \left( D_{\alpha }D^{\alpha
}\phi \right) ^{2}-\left( D^{\mu }D^{\nu }\phi \right) \left( D_{\mu
}D_{_{\nu }}\phi \right) \right]  \notag \\
\mathcal{L}_{5}\left[ G_{5}\right] &\equiv &G_{5}\left( \phi ,X\right) 
\tilde{G}_{\mu \upsilon }D^{\mu }D^{\nu }\phi  \notag \\
&&+\frac{1}{3}G_{5}\left( \phi ,X\right) \left[ \left( D_{\mu }D^{\mu }\phi
\right) ^{3}-3\left( D_{\mu }D^{\mu }\phi \right) \left( D^{\gamma }D^{\nu
}\phi \right) \left( D_{\gamma }D_{\nu }\phi \right) \right]  \notag \\
&&+2\left( D_{\mu }D_{\nu }\phi \right) \left( D^{\sigma }D^{\nu }\phi
\right) \left( D_{\sigma }D^{\mu }\phi \right)  \label{a3-1}
\end{eqnarray}

\section{\textbf{Appendix 3: Lagrangian terms \protect\ref{l11} in tensor
language}}

We consider the tensor form of the eight terms of the Lagrangian
corresponding to teh action (\ref{l11}). To do this we will use the equation
(\ref{h0'}), with $\tilde{k}_{m}=\tilde{e}_{m}^{\text{ \ }\mu }\partial
_{\mu }\phi $.

\textbf{First term:} in this case it is direct to see that 
\begin{equation}
A\tilde{\epsilon}_{mnpq}\tilde{R}^{mn}\tilde{e}^{p}\tilde{e}^{q}=4A\sqrt{-%
\tilde{g}}\tilde{R}d^{4}\tilde{x}.  \label{a2-1}
\end{equation}

\textbf{Second term:} writing the second term in the form,

\begin{eqnarray}
A\tilde{\epsilon}_{mnpq}\tilde{k}_{\ f}^{m}\tilde{k}^{fn}\tilde{e}^{p}\tilde{%
e}^{q} &=&\frac{A}{9}\tilde{\epsilon}_{mnpq}\left( \tilde{k}^{m}\tilde{e}%
_{f}-\tilde{k}_{f}\tilde{e}^{m}\right) \left( \tilde{k}^{f}\tilde{e}^{n}-%
\tilde{k}^{n}\tilde{e}^{f}\right) \tilde{e}^{p}\tilde{e}^{q}  \notag \\
&=&\frac{A}{9}\tilde{\epsilon}_{mnpq}\left( \tilde{k}^{m}\tilde{e}_{f}\tilde{%
k}^{f}\tilde{e}^{n}-\tilde{k}_{f}\tilde{k}^{f}\tilde{e}^{m}\tilde{e}^{n}-%
\tilde{k}^{m}\tilde{k}^{n}\tilde{e}_{f}\tilde{e}^{f}\right.  \notag \\
&&\left. -\tilde{k}_{f}\tilde{e}^{f}\tilde{e}^{m}\tilde{k}^{n}\right) \tilde{%
e}^{p}\tilde{e}^{q},  \label{a2-2}
\end{eqnarray}%
and considering that $\tilde{e}^{f}\tilde{k}_{f}=\left( \tilde{e}_{\text{ }%
\nu }^{f\text{ }}dx^{\nu }\right) \left( \tilde{e}_{f}^{\text{ \ }\mu
}\partial _{\mu }\phi \right) =\delta _{\nu }^{\text{ }\mu }dx^{\nu
}\partial _{\mu }\phi =d\phi $, we have 
\begin{equation}
A\tilde{\epsilon}_{mnpq}\tilde{k}_{\ f}^{m}\tilde{k}^{fn}\tilde{e}^{p}\tilde{%
e}^{q}=\frac{2A}{9}\tilde{\epsilon}_{mnpq}d\varphi \tilde{k}^{m}\tilde{e}^{n}%
\tilde{e}^{p}\tilde{e}^{q}-\frac{A}{9}\tilde{\epsilon}_{mnpq}\tilde{k}^{2}%
\tilde{e}^{m}\tilde{e}^{n}\tilde{e}^{p}\tilde{e}^{q},  \label{a2-3}
\end{equation}%
where,

\begin{eqnarray}
\tilde{\epsilon}_{mnpq}\tilde{k}^{2}\tilde{e}^{m}\tilde{e}^{n}\tilde{e}^{p}%
\tilde{e}^{q} &=&\tilde{\epsilon}_{mnpq}e_{\text{ }\alpha }^{m}e_{\text{ }%
\beta }^{n}\tilde{e}_{\text{ }\gamma }^{p}\tilde{e}_{\text{ }\delta
}^{q}dx^{\alpha }dx^{\beta }dx^{\gamma }dx^{\delta }D_{\mu }\varphi D^{\mu
}\varphi  \notag \\
&=&\sqrt{\tilde{g}}d^{4}x\tilde{\epsilon}_{\alpha \beta \gamma \delta }%
\tilde{\epsilon}^{\alpha \beta \gamma \delta }D_{\mu }\varphi D^{\mu
}\varphi =\sqrt{\tilde{g}}D_{\mu }\varphi D^{\mu }\varphi 4!d^{4}x  \notag \\
2\tilde{\epsilon}_{mnpq}d\varphi \tilde{k}^{m}\tilde{e}^{n}\tilde{e}^{p}%
\tilde{e}^{q} &=&2\tilde{\epsilon}_{mnpq}e_{\text{ }\alpha }^{m}e_{\text{ }%
\beta }^{n}\tilde{e}_{\text{ }\gamma }^{p}\tilde{e}_{\text{ }\delta
}^{q}dx^{\mu }dx^{\beta }dx^{\gamma }dx^{\delta }\partial _{\mu }\varphi
D^{\alpha }\varphi  \notag \\
&=&2\sqrt{\tilde{g}}d^{4}x\tilde{\epsilon}_{\alpha \beta \gamma \delta }%
\tilde{\epsilon}^{\mu \beta \gamma \delta }D_{\mu }\varphi D^{\alpha }\varphi
\notag \\
&=&2\cdot 3!D_{\mu }\varphi D^{\mu }\varphi \sqrt{\tilde{g}}d^{4}x
\label{a2-4}
\end{eqnarray}%
so that, 
\begin{eqnarray}
A\tilde{\epsilon}_{mnpq}\tilde{k}_{\ f}^{m}\tilde{k}^{fn}\tilde{e}^{p}\tilde{%
e}^{q} &=&\frac{A}{9}\left( 2\cdot 3!D_{\mu }\varphi D^{\mu }\varphi
-4!D_{\mu }\varphi D^{\mu }\varphi \right) \sqrt{\tilde{g}}d^{4}x  \notag \\
&=&-\frac{4}{3}AD_{\mu }\varphi D^{\mu }\varphi \sqrt{\tilde{g}}d^{4}x
\label{a2-5}
\end{eqnarray}

\textbf{Third term:} for the third term, we have

\begin{equation}
A\tilde{\epsilon}_{mnpq}D\tilde{k}^{mn}\tilde{e}^{p}\tilde{e}^{q}=-\frac{2}{3%
}A\tilde{\epsilon}_{mnpq}D\tilde{k}^{m}\tilde{e}^{n}\tilde{e}^{p}\tilde{e}%
^{q},  \label{a2-6}
\end{equation}%
where we have taken into account that the absence of four-dimensional
torsion, that is, $\tilde{T}^{m}=D\tilde{e}^{m}=0$. In tensor language (\ref%
{a2-6}) takes the form,

\begin{eqnarray}
A\tilde{\epsilon}_{mnpq}D\tilde{k}^{mn}\tilde{e}^{p}\tilde{e}^{q} &=&-\frac{2%
}{3}A\tilde{\epsilon}_{mnpq}e_{\text{ }\alpha }^{m}\tilde{e}_{\text{ }\beta
}^{n}\tilde{e}_{\text{ }\gamma }^{p}\tilde{e}_{\text{ }\delta }^{q}D_{\mu
}D^{\alpha }\varphi dx^{\mu }dx^{\beta }dx^{\gamma }dx^{\delta }  \notag \\
&=&-\frac{2}{3}A\tilde{\epsilon}_{\alpha \beta \gamma \delta }\tilde{\epsilon%
}^{\mu \beta \gamma \delta }D_{\mu }D^{\alpha }\varphi \sqrt{-\tilde{g}}d^{4}%
\tilde{x}  \notag \\
&=&-4AD_{\mu }D^{\mu }\varphi \sqrt{-\tilde{g}}d^{4}\tilde{x}.  \label{a2-7}
\end{eqnarray}

\textbf{Fourth term:} in this case it is direct to see that

\begin{equation}
B\tilde{\epsilon}_{mnpq}\tilde{e}^{m}\tilde{e}^{n}\tilde{e}^{p}\tilde{e}%
^{q}=24B\sqrt{-\tilde{g}}d^{4}\tilde{x}.  \label{a2-8}
\end{equation}

Fifth term: in this case we can write

\begin{eqnarray}
C\tilde{\varepsilon}_{mnpq}D\tilde{k}^{mn}D\tilde{k}^{pq} &=&\frac{1}{9}C%
\tilde{\varepsilon}_{mnpq}\left( D\tilde{k}^{m}\tilde{e}^{n}-D\tilde{k}^{n}%
\tilde{e}^{m}\right) \left( D\tilde{k}^{p}\tilde{e}^{q}-D\tilde{k}^{q}\tilde{%
e}^{p}\right)  \notag \\
&&\frac{4}{9}C\tilde{\varepsilon}_{mnpq}D\tilde{k}^{m}\tilde{e}^{n}D\tilde{k}%
^{p}\tilde{e}^{q},
\end{eqnarray}%
where we have taken into account the absence of four-dimensional torsion. So
we have,

\begin{eqnarray}
C\tilde{\varepsilon}_{mnpq}D\tilde{k}^{mn}D\tilde{k}^{pq} &=&\frac{4C}{9}%
\tilde{\varepsilon}_{mnpq}\tilde{e}_{\alpha }^{m}\tilde{e}_{\beta }^{n}%
\tilde{e}_{\nu }^{p}\tilde{e}_{\delta }^{q}dx^{\mu }dx^{\beta }dx^{\gamma
}dx^{\delta }D_{\mu }D^{\alpha }\varphi D_{\gamma }D^{\nu }\varphi  \notag \\
&=&\frac{4C}{9}\sqrt{-\tilde{g}}d^{4}\tilde{x}\tilde{\varepsilon}_{\alpha
\beta \gamma \delta }\varepsilon ^{\mu \beta \nu \delta }D_{\mu }D^{\alpha
}\varphi D_{\nu }D^{\gamma }\varphi  \notag \\
&=&\frac{8C}{9}\left\{ \left( D_{\alpha }D^{\alpha }\varphi \right)
^{2}-D_{\gamma }D^{\nu }\varphi D_{\nu }D^{\gamma }\varphi \right\} \sqrt{-%
\tilde{g}}d^{4}\tilde{x}  \label{a2-9}
\end{eqnarray}

\textbf{Sixth term:} using the previous result,%
\begin{equation*}
\tilde{k}_{\ f}^{m}\tilde{k}^{fn}=\frac{1}{3^{2}}\left( 2d\varphi \tilde{k}%
^{m}\tilde{e}^{n}-\tilde{k}^{2}\tilde{e}^{m}\tilde{e}^{n}\right) ,
\end{equation*}%
we have, 
\begin{eqnarray}
C\tilde{\varepsilon}_{mnpq}\tilde{k}_{\ f}^{m}\tilde{k}^{fn}\tilde{k}_{\
g}^{p}\tilde{k}^{gq} &=&\frac{C}{3^{4}}\tilde{\varepsilon}_{mnpq}\left(
2d\varphi \tilde{k}^{m}\tilde{e}^{n}-\tilde{k}^{2}\tilde{e}^{m}\tilde{e}%
^{n}\right) \left( 2d\varphi \tilde{k}^{p}\tilde{e}^{q}-\tilde{k}^{2}\tilde{e%
}^{p}\tilde{e}^{q}\right)  \notag \\
&=&\frac{C}{3^{4}}\tilde{\varepsilon}_{mnpq}\left\{ 4\left( d\varphi \right)
^{2}\tilde{k}^{m}\tilde{e}^{n}\tilde{k}^{p}\tilde{e}^{q}-4d\varphi \tilde{k}%
^{2}\tilde{k}^{m}\tilde{e}^{n}\tilde{e}^{p}\tilde{e}^{q}\right.  \notag \\
&&\left. +\tilde{k}^{4}\tilde{e}^{m}\tilde{e}^{n}\tilde{e}^{p}\tilde{e}%
^{q}\right\} .  \label{a2-10}
\end{eqnarray}

To obtain this result in tensor language, let's analyze each term
separately. Indeed,%
\begin{eqnarray}
\frac{4C}{3^{4}}\tilde{\varepsilon}_{mnpq}\left( d\varphi \right) ^{2}\tilde{%
k}^{m}\tilde{e}^{n}\tilde{k}^{p}\tilde{e}^{q} &=&\frac{4C}{3^{4}}\tilde{%
\varepsilon}_{mnpq}\tilde{e}_{\alpha }^{m}\tilde{e}_{\beta }^{n}\tilde{e}%
_{\gamma }^{p}\tilde{e}_{\delta }^{q}dx^{\mu }dx^{\nu }dx^{\beta }dx^{\delta
}\partial _{\mu }\varphi \partial _{\nu }\phi D^{\alpha }\varphi D^{\gamma
}\varphi  \notag \\
&=&-\frac{4C}{3^{4}}\sqrt{-\tilde{g}}d^{4}\tilde{x}\tilde{\varepsilon}%
_{\alpha \gamma \beta \delta }\tilde{\varepsilon}^{\mu \nu \beta \delta
}\partial _{\mu }\varphi \partial _{\nu }\varphi D^{\alpha }\varphi
D^{\gamma }\varphi  \notag \\
&=&-\frac{8C}{3^{4}}\sqrt{-\tilde{g}}d^{4}\tilde{x}\left( D_{\alpha }\varphi
D^{\alpha }\varphi D_{\gamma }\varphi D^{\gamma }\varphi \right.  \notag \\
\left. -D_{\alpha }\varphi D^{\alpha }\varphi D_{\gamma }\varphi D^{\gamma
}\varphi \right) &=&0  \label{a2-11}
\end{eqnarray}

\begin{eqnarray}
-\frac{4C}{3^{4}}\tilde{\varepsilon}_{mnpq}d\varphi \tilde{k}^{2}\tilde{k}%
^{m}\tilde{e}^{n}\tilde{e}^{p}\tilde{e}^{q} &=&-\frac{4C}{3^{4}}\tilde{%
\varepsilon}_{mnpq}\tilde{e}_{\text{ }\alpha }^{m}\tilde{e}_{\text{ }\beta
}^{n}\tilde{e}_{\text{ }\gamma }^{p}\tilde{e}_{\text{ }\delta }^{q}dx^{\mu
}dx^{\beta }dx^{\gamma }dx^{\delta }\partial _{\mu }\varphi D^{\alpha
}\varphi \tilde{k}^{2}  \notag \\
&=&-\frac{4C}{3^{4}}\sqrt{-\tilde{g}}d^{4}\tilde{x}\tilde{\varepsilon}%
_{\alpha \beta \gamma \delta }\tilde{\varepsilon}^{\mu \beta \gamma \delta
}D_{\mu }\varphi D^{\alpha }\varphi \tilde{k}^{2}  \notag \\
&=&-\frac{8C}{3^{3}}\sqrt{-\tilde{g}}d^{4}\tilde{x}D_{\alpha }\varphi
D^{\alpha }\varphi D_{\nu }\varphi D^{\nu }\varphi  \notag \\
&=&-\frac{8C}{3^{3}}\sqrt{-\tilde{g}}d^{4}\tilde{x}\left( D_{\alpha }\varphi
\right) ^{4}  \label{a2-12}
\end{eqnarray}%
\begin{eqnarray}
\frac{C}{3^{4}}\tilde{\varepsilon}_{mnpq}\tilde{k}^{4}\tilde{e}^{m}\tilde{e}%
^{n}\tilde{e}^{p}\tilde{e}^{q} &=&\frac{C}{3^{4}}\tilde{\varepsilon}_{mnpq}%
\tilde{e}_{\text{ }\alpha }^{m}\tilde{e}_{\text{ }\beta }^{n}\tilde{e}_{%
\text{ }\gamma }^{p}\tilde{e}_{\text{ }\delta }^{q}dx^{\alpha }dx^{\beta
}dx^{\gamma }dx^{\delta }D_{\mu }\varphi D^{\mu }\varphi D_{\nu }\varphi
D^{\nu }\varphi  \notag \\
&=&\frac{C}{3^{4}}\sqrt{-\tilde{g}}d^{4}\tilde{x}\tilde{\varepsilon}_{\alpha
\beta \gamma \delta }\tilde{\varepsilon}^{\alpha \beta \gamma \delta }D_{\mu
}\varphi D^{\mu }\varphi D_{\nu }\varphi D^{\nu }\varphi  \notag \\
&=&\frac{8C}{3^{3}}\sqrt{-\tilde{g}}d^{4}\tilde{x}\left( D_{\mu }\varphi
\right) ^{4}.  \label{a2-13}
\end{eqnarray}

Introducing (\ref{a2-11}, \ref{a2-12}, \ref{a2-13}) into (\ref{a2-11}), we
find%
\begin{equation}
C\tilde{\varepsilon}_{mnpq}\tilde{k}_{\ f}^{m}\tilde{k}^{fn}\tilde{k}_{\
g}^{p}\tilde{k}^{gq}=0,  \label{a2-14}
\end{equation}%
which proves that the sixth term is null.

\textbf{Seventh term:} In this case we can write, 
\begin{eqnarray}
2C\tilde{\varepsilon}_{mnpq}\tilde{R}^{mn}\tilde{k}_{\ f}^{p}\tilde{k}^{fq}
&=&\frac{2C}{3^{2}}\tilde{\varepsilon}_{mnpq}\tilde{R}^{mn}\left( 2d\varphi 
\tilde{k}^{p}\tilde{e}^{q}-\tilde{k}^{2}\tilde{e}^{p}\tilde{e}^{q}\right) 
\notag \\
&=&\frac{4C}{3^{2}}d\varphi \tilde{\varepsilon}_{mnpq}\tilde{R}^{mn}\tilde{k}%
^{p}\tilde{e}^{q}  \notag \\
&&-\frac{2C}{3^{2}}\tilde{k}^{2}\tilde{\varepsilon}_{mnpq}\tilde{R}^{mn}%
\tilde{e}^{p}\tilde{e}^{q},  \label{a2-15}
\end{eqnarray}%
where, 
\begin{eqnarray}
\frac{4C}{3^{2}}d\phi \tilde{\varepsilon}_{mnpq}\tilde{R}^{mn}\tilde{k}^{p}%
\tilde{e}^{q} &=&\frac{2C}{3^{2}}D_{\mu }\phi \tilde{\varepsilon}_{mnpq}%
\tilde{e}_{\text{ }\alpha }^{m}\tilde{e}_{\text{ }\beta }^{n}\tilde{e}_{%
\text{ }\gamma }^{p}\tilde{e}_{\text{ }\delta }^{q}dx^{\mu }dx^{\lambda
}dx^{\rho }dx^{\delta }\tilde{R}_{\lambda \rho }^{\alpha \beta }D^{\gamma
}\varphi  \notag \\
&=&\frac{2C}{3^{2}}\sqrt{-\tilde{g}}d^{4}\tilde{x}\tilde{\varepsilon}%
_{\alpha \beta \gamma \delta }\tilde{\varepsilon}^{\mu \lambda \rho \delta
}D_{\mu }\varphi D^{\gamma }\varphi \tilde{R}_{\lambda \rho }^{\alpha \beta }
\notag \\
&=&\frac{2C}{3^{2}}\sqrt{-\tilde{g}}d^{4}\tilde{x}\delta _{\alpha \beta
\gamma }^{\mu \lambda \rho }D_{\mu }\varphi D^{\gamma }\varphi \tilde{R}%
_{\lambda \rho }^{\alpha \beta },  \notag \\
&=&\frac{-8C}{3^{2}}\sqrt{-\tilde{g}}d^{4}\tilde{x}D^{\alpha }\varphi
D^{\beta }\varphi \left\{ \tilde{R}_{\alpha \beta }-\frac{1}{2}g_{\alpha
\beta }\tilde{R}\right\}  \label{a2-16}
\end{eqnarray}

\begin{equation}
-\frac{2C}{3^{2}}\tilde{k}^{2}\tilde{\varepsilon}_{mnpq}\tilde{R}^{mn}\tilde{%
e}^{p}\tilde{e}^{q}=-\frac{8C}{3^{2}}\sqrt{-\tilde{g}}d^{4}\tilde{x}D_{\nu
}\varphi D^{\nu }\varphi \tilde{R}.  \label{a2-17}
\end{equation}%
Introducing (\ref{a2-16}, \ref{a2-17}) in (\ref{a2-15}),we have%
\begin{equation}
2C\tilde{\varepsilon}_{mnpq}\tilde{R}^{mn}\tilde{k}_{\ f}^{p}\tilde{k}^{fq}=%
\frac{-8C}{3^{2}}\sqrt{-\tilde{g}}d^{4}\tilde{x}D^{\alpha }\varphi D^{\beta
}\varphi \tilde{G}_{\alpha \beta }-\frac{8C}{3^{2}}\sqrt{-\tilde{g}}d^{4}%
\tilde{x}D_{\nu }\varphi D^{\nu }\varphi \tilde{R}  \label{a2-18}
\end{equation}

\textbf{Eighth term:} as in the previous cases we write

\begin{eqnarray}
2C\tilde{\varepsilon}_{mnpq}D\tilde{k}^{mn}\tilde{k}_{\ f}^{p}\tilde{k}^{fq}
&=&-\frac{2}{3\cdot 3^{2}}C\tilde{\varepsilon}_{mnpq}D\left( \tilde{k}^{m}%
\tilde{e}^{n}-\tilde{k}^{n}\tilde{e}^{m}\right) \left( 2d\phi \tilde{k}^{p}%
\tilde{e}^{q}-\tilde{k}^{2}\tilde{e}^{p}\tilde{e}^{q}\right)   \notag \\
&=&-\frac{8C}{3^{3}}d\phi \tilde{\varepsilon}_{mnpq}D\tilde{k}^{m}\tilde{e}%
^{n}\tilde{k}^{p}\tilde{e}^{q}  \notag \\
&&+\frac{4C}{3^{3}}\tilde{k}^{2}\tilde{\varepsilon}_{mnpq}D\tilde{k}^{m}%
\tilde{e}^{n}\tilde{e}^{p}\tilde{e}^{q}.  \label{a2-19}
\end{eqnarray}

Analyzing each term separately, we have,%
\begin{eqnarray}
-\frac{8C}{3^{3}}d\varphi \tilde{\varepsilon}_{mnpq}D\tilde{k}^{m}\tilde{e}%
^{n}\tilde{k}^{p}\tilde{e}^{q} &=&-\frac{8C}{3^{3}}\tilde{\varepsilon}_{mnpq}%
\tilde{e}_{\text{ }\alpha }^{m}\tilde{e}_{\text{ }\beta }^{n}\tilde{e}_{%
\text{ }\gamma }^{p}\tilde{e}_{\text{ }\delta }^{q}dx^{\mu }dx^{\nu
}dx^{\beta }dx^{\delta }D_{\mu }\varphi D_{\nu }D^{\alpha }\varphi D^{\gamma
}\varphi  \notag \\
&=&\frac{8C}{3^{3}}\sqrt{-\tilde{g}}d^{4}\tilde{x}\tilde{\varepsilon}%
_{\alpha \gamma \beta \delta }\tilde{\varepsilon}^{\mu \nu \beta \delta
}D_{\mu }\varphi D_{\nu }D^{\alpha }\varphi D^{\gamma }\varphi  \notag \\
&=&\frac{16C}{3^{3}}\sqrt{-\tilde{g}}d^{4}\tilde{x}\left\{ D_{\alpha
}\varphi \left( D_{\gamma }D^{\alpha }\varphi \right) D^{\gamma }\varphi
-D_{\gamma }\varphi D^{\gamma }\varphi \left( D_{\alpha }D^{\alpha }\varphi
\right) \right\}  \notag \\
&&\frac{16C}{3^{3}}\sqrt{-\tilde{g}}d^{4}\tilde{x}\left\{ \frac{1}{2}%
D_{\gamma }\left[ D_{\alpha }\varphi D^{\alpha }\varphi D^{\gamma }\varphi %
\right] \right.  \notag \\
&&\left. -\frac{3}{2}D_{\alpha }\varphi D^{\alpha }\varphi \left( D_{\gamma
}D^{\gamma }\varphi \right) \right\} ,  \label{a2-20}
\end{eqnarray}%
where we have used 
\begin{equation*}
\left( D_{\gamma }D_{\alpha }\varphi \right) D^{\alpha }\varphi D^{\gamma
}\varphi =\frac{1}{2}D_{\gamma }\left[ D_{\alpha }\varphi D^{\alpha }\varphi
D^{\gamma }\varphi \right] -\frac{1}{2}D_{\alpha }\varphi D^{\alpha }\varphi
D_{\gamma }D^{\gamma }\varphi .
\end{equation*}

For the second term of (\ref{a2-19}), it is found,%
\begin{eqnarray}
\frac{4C}{3^{3}}\tilde{k}^{2}\tilde{\varepsilon}_{mnpq}D\tilde{k}^{m}\tilde{e%
}^{n}\tilde{e}^{p}\tilde{e}^{q} &=&\frac{4C}{3^{3}}\tilde{\varepsilon}_{mnpq}%
\tilde{e}_{\text{ }\alpha }^{m}\tilde{e}_{\beta }^{n}\tilde{e}_{\gamma }^{p}%
\tilde{e}_{\delta }^{q}dx^{\nu }dx^{\beta }dx^{\gamma }dx^{\delta }D_{\mu
}\varphi D^{\mu }\varphi D_{\nu }D^{\alpha }\varphi  \notag \\
&=&\frac{4C}{3^{3}}\sqrt{-\tilde{g}}d^{4}\tilde{x}\tilde{\varepsilon}%
_{\alpha \beta \gamma \delta }\tilde{\varepsilon}^{\nu \beta \gamma \delta
}D_{\mu }\varphi D^{\mu }\varphi D_{\nu }D^{\alpha }\varphi  \notag \\
&=&\frac{8C}{3^{2}}\sqrt{-\tilde{g}}d^{4}\tilde{x}D_{\mu }\varphi D^{\mu
}\varphi D_{\alpha }D^{\alpha }\varphi .  \label{a2-21}
\end{eqnarray}

Introducing (\ref{a2-20}, \ref{a2-21}) into (\ref{a2-19}), we find that the
eighth term is given by,%
\begin{eqnarray}
2C\tilde{\varepsilon}_{mnpq}D\tilde{k}^{mn}\tilde{k}_{\ f}^{p}\tilde{k}^{fq}
&=&\frac{16C}{3^{3}}\sqrt{-\tilde{g}}d^{4}\tilde{x}\left\{ \frac{1}{2}%
D_{\gamma }\left[ D_{\alpha }\varphi D^{\alpha }\varphi D^{\gamma }\varphi %
\right] -\frac{3}{2}D_{\alpha }\varphi D^{\alpha }\varphi \left( D_{\gamma
}D^{\gamma }\varphi \right) \right\}  \notag \\
&=&\sqrt{-\tilde{g}}d^{4}\tilde{x}\frac{8C}{3^{3}}D_{\gamma }A^{\gamma },%
\text{ con \ \ }A^{\gamma }=D_{\alpha }\varphi D^{\alpha }\varphi D^{\gamma
}\varphi  \label{a2-22}
\end{eqnarray}

\textbf{Acknowledgements: \ }The authors wish to thanks Stephanie Caro,
Sebastian Salgado, Cristian Vera, for enlightening discussions. P. S. was
supported by Fondo Nacional de Desarrollo Cient\'{\i}fico y Tecnol\'{o}gico
(FONDECYT, Chile) through Grants \#1180681 and \#1211219. DMP acknowledges
financial support from the Chilean government through Fondecyt grants Grant
\#11240533. L.A. was supported by Fondecyt grant \#3220805. VCO was
supported by Universidad de Concepci\'{o}n\textit{,} Chile and JD was
supported by Universidad Arturo Prat\textit{,} Chile.


\begin{thebibliography}{99}
\bibitem{exp4} An expansion is, in general, an algebra dimension-changing
process, i.e., is a way to obtain new algebras of increasingly higher
dimensions from a given one. A physical motivation for increasing the
dimension of Lie algebras is that increasing the number of generators of an
algebra is a non-trivial way of enlarging spacetime symmetries. Examples of
this can be found in Refs. \cite{bonanos,gomis}, where applications of
Maxwell's algebra in gravity were studied (This algebra, also known as $%
\mathfrak{B}_{4}$\ algebra, is a modification to the Poincar\'{e} symmetries
and can be obtained, via S-expansion, from the anti-de Sitter (AdS)
algebra). Another interesting modification to the Poincar\'{e} symmetries
are the so-called generalized Poincar\'{e} algebras \cite{ss} of which the $%
\mathcal{M\ }$algebra is an example. \ \ \
\end{thebibliography}
\end{document}